\newcount\Comments  
\documentclass[twocolumn]{aastex631}
\usepackage{hyperref}
\usepackage{soul}
\usepackage{xcolor}
\usepackage{graphicx}
\usepackage{parskip}
\usepackage{amsmath}
\usepackage{supertabular}
\usepackage{booktabs}

\definecolor{purple}{rgb}{0.6,0,1}
\definecolor{blue2}{rgb}{0,.5,1}
\newcommand{\kibitz}[2]{\ifnum\Comments=1\textcolor{#1}{#2}\fi}

\newcommand{\lsim}{\mbox{\rlap{\hbox{\lower2pt\hbox{\ensuremath{\sim}}}}\raise2pt\hbox{\ensuremath{<}}}}%
\newcommand{\gsim}{\mbox{\rlap{\hbox{\lower2pt\hbox{\ensuremath{\sim}}}}\raise2pt\hbox{\ensuremath{>}}}}%
\renewcommand{\gtrsim}  {\ensuremath{\gsim}}
\renewcommand{\lesssim} {\ensuremath{\lsim}}
\newcommand\cge{{\,$\gtrsim$\,}}%
\newcommand\cle{{\,$\lesssim$\,}}%
\newcommand\mAB{\ensuremath{m_{\rm AB}}}%
\newcommand{\degree}{\ensuremath{^{\circ}}}%

\newcommand{\HST}{\emph{HST}}%
\newcommand{\DELETED}[1]{\relax}%
{\relax}%

\shortauthors{Kramer et al.}

\begin{document}

\title{SKYSURF-3: Testing Crowded Object Catalogs in the Hubble eXtreme Deep Field Mosaics to Study Sample Incompleteness from an Extragalactic Background Light Perspective}

\author[0000-0003-0238-8806]{Darby M. Kramer}
\affiliation{School of Earth and Space Exploration,
Arizona State University, Tempe, AZ 85287 USA}

\author[0000-0001-6650-2853]{Timothy Carleton}
\affiliation{School of Earth and Space Exploration,
Arizona State University, Tempe, AZ 85287 USA}

\author[0000-0003-3329-1337]{Seth. H. Cohen}
\affiliation{School of Earth and Space Exploration,
Arizona State University, Tempe, AZ 85287 USA}

\author[0000-0003-1268-5230]{Rolf A. Jansen}
\affiliation{School of Earth and Space Exploration,
Arizona State University, Tempe, AZ 85287 USA}

\author[0000-0001-8156-6281]{Rogier A. Windhorst}
\affiliation{School of Earth and Space Exploration,
Arizona State University, Tempe, AZ 85287 USA}

\author[0000-0001-9440-8872]{Norman A. Grogin}
\affiliation{Space Telescope Science Institute, 3700 San Martin Dr., Baltimore, MD 21218, USA}

\author[0000-0002-6610-2048]{Anton M. Koekemoer}
\affiliation{Space Telescope Science Institute, 3700 San Martin Dr., Baltimore, MD 21218, USA}

\author[0000-0001-6529-8416]{John W. Mackenty}
\affiliation{Space Telescope Science Institute, 3700 San Martin Dr., Baltimore, MD 21218, USA}

\author[0000-0003-3382-5941]{Norbert Pirzkal}
\affiliation{Space Telescope Science Institute, 3700 San Martin Dr., Baltimore, MD 21218, USA}

\begin{abstract}
Extragalactic Background Light (EBL) studies have revealed a significant discrepancy between direct measurements --- via instruments measuring “bare” sky from which Zodiacal and Galactic light models are subtracted --- and measurements of the Integrated Galaxy Light (IGL). This discrepancy could lie in either method, whether it be an incomplete Zodiacal model or missed faint galaxies in the IGL calculations. It has been proposed that the discrepancy is due to deep galaxy surveys, such as those with the \emph{Hubble Space Telescope} (\HST), missing up to half of the faint galaxies with 24\cle\mAB\cle29 mag. We address this possibility by simulating higher number densities of galaxies, and so assess incompleteness due to object overlap, with three replications of the Hubble UltraDeep Field (HUDF). \texttt{SourceExtractor} is used to compare the recovered counts and photometry to the original HUDF, allowing us to assess how many galaxies may have been missed due to confusion, i.e., due to blending with neighboring faint galaxies. This exercise reveals that, while up to 50\% of faint galaxies with 28\cle\mAB\cle29 mag were missed or blended with neighboring objects in certain filters, not enough were missed to account for the EBL discrepancy alone in any of the replications.
\end{abstract}

\keywords{Cosmology: Extragalactic Background Light --- Galaxies: Galaxy Counts --- Solar System: Zodiacal Light --- Instrument: Hubble Space Telescope}

\section{Introduction}\label{sec:intro}

Extragalactic Background Light (EBL) is the dominant background in the Universe after the cosmic microwave background (CMB), and is generally considered to be comprised of the short wavelength (0.1\,\micron) ultraviolet to long wavelength (1000\,\micron) far-infrared emission \citep{Driver_2016,mandv2019}, though this depends on convention, and can also be considered to include the entire electromagnetic spectrum. In terms of energy density, two main components make up the EBL: the cosmic optical background (COB) and the cosmic infrared background (CIB). Because a range of astrophysical processes emit significant amounts of photons at ultraviolet (UV), optical and infrared (IR) wavelengths, the EBL includes radiation from several types of objects in the Universe such as stars, AGN, and dust \citep{madaupozzetti2000}. There are two main ways that EBL measurements are obtained: (A) direct measurements of the sky, from which Zodiacal and diffuse Galactic light is subtracted, and (B) integrated galactic light (IGL) from galaxy surveys. Both methods are demonstrated in, e.g., \citet{bernstein2007}. The EBL is expected to be highly dominated by galaxy light, and current technology allows for deep images at UV, optical, and IR wavelengths. Thus, integrating the galaxy light in deep sky catalogs should yield an accurate representation of the EBL, unless a significant diffuse EBL component is not present in the discrete galaxy catalogs. 

The reason this is still an active area of study is that, although the IGL at IR wavelengths is in fairly good agreement with direct CIB measurements, the UV and optical IGL still do not agree with the directly-estimated diffuse COB: current direct measurements of the EBL are significantly higher than the IGL calculations and COB models \citep{Keenan_2010,Driver_2016,hill2018}. Uncertainties in the direct EBL measurements come from foregrounds such as Zodiacal light and dust from the Milky Way \citep{madaupozzetti2000,matsumoto2015}, while uncertainties in IGL calculations come from instrumental limitations and the number density of galaxies on the sky. 

\citet{driver2021measuring} addresses the issue with high energy gamma ray studies, such as HESS and MAGIC, which get information from blazar photons that interact with the EBL on their path toward Earth. The blazar spectra are expected to have absorption features at the wavelengths where the gamma ray and EBL photons interact, constraining the permitted levels of the EBL intensity with a $\sim$50\% error range \citep{hess2006,magic2019}. These predictions match the IGL calculations and EBL models well \citep{VERITAS}, providing independent constraints that suggest the higher direct EBL measurements may still include components that need to be subtracted. Deep space missions, such as Pioneer 10/11 and New Horizons, also improve direct EBL measurements because they have much less Zodiacal light to contend with than telescopes closer to the Sun \citep{pioneer11,Lauer2021,Lauer_2022}. These missions have provided some direct EBL data points which are lower than those from Low Earth Orbit or ground-based experiments \citep[see, e.g., ][]{bernstein2007} Even with these improved constraints, the cause of the EBL discrepancy is still unidentified \citep{driver2021measuring}. Among the possibilities are: (A) a Zodiacal component that is missing from or underestimated in current models, causing an overestimation of the EBL in direct measurements; (B) missed faint diffuse galaxies or extended galaxy outskirts in galaxy surveys, causing an underestimation of the IGL; or (C) both.

\citet{Lauer2021} propose that their extra, unexplained diffuse flux component of the COB might result if galaxy surveys were missing up to half of all the faint galaxies that exist. If these faint galaxies are being missed, they are not counted in the IGL calculations and could cause the true EBL to be underestimated. Their paper explains that the faint-end slope  could drop off sharply at \mAB\ $>$ 24 mag, i.e., that surveys miss significant numbers of faint galaxies with \mAB\cle30 mag. Such galaxies are very faint, however, and contribute much less than 25\% of the IGL according to \citet{rogier2022}. That paper and \citet{tim2022} suggest that one would need a factor of 4--8 times more faint galaxies with 24\cle\mAB\cle30 mag to make up for the missing diffuse EBL flux \citep{Lauer2021, Lauer_2022, bernstein2007, hill2018}.

Deep extragalactic surveys with \HST, starting with the Hubble Deep Field \citep{1996AJ....112.1335W}, and subsequently the Hubble UltraDeep Field/eXtreme Deep Field (HUDF/XDF) \citep{2006AJ....132.1729B, 2010ApJ...709L..16O, Ellis2013, Koekemoer2013, 2013ApJS..209....6I, rafelski2015} have detected and resolved distant galaxies, with some candidates above z $\sim$ 10 and to apparent magnitudes \mAB\ $\sim$ 30 mag, constraining the observed galaxy luminosity function \citep{Oesch&Bouwens2009,Finkelstein2012,2011ApJS..193...27W, windhorst2021}. Even though new galaxy surveys continue to reach deeper and wider on the sky, their galaxy counts are still complicated by instrumental effects and object confusion. Instrumental confusion is the inability of the optical system of telescope, camera, and detector to resolve and separate faint objects because of limits that depend on the image point-source sensitivity. On the contrary, natural confusion is the inability to resolve faint, extended objects due to their surface brightness and statistical overlap. Any faint object that is even slightly resolved can overlap other objects. This causes incompleteness at dim magnitudes.

Although instrumental confusion can be mitigated with larger apertures, natural confusion is a fundamental problem that will persist as galaxy surveys reach deeper. Confusion is thus a major source of noise in the deep images being acquired today \citep{herschelconfusion}. This work focuses on generating and studying the effects of natural confusion alone, as it has stronger limits on present-day surveys. Deep surveys with \HST\ run into the natural confusion limit of 1 object per 25 -- 50 independent beams, where the beam is the average object diameter at a given flux \citep[][their Figure 3]{Serjeant_1997, Silva_2004, Windhorst2008}.

\begin{figure*}[h!]
\centering
    \includegraphics[width=0.58\textwidth]{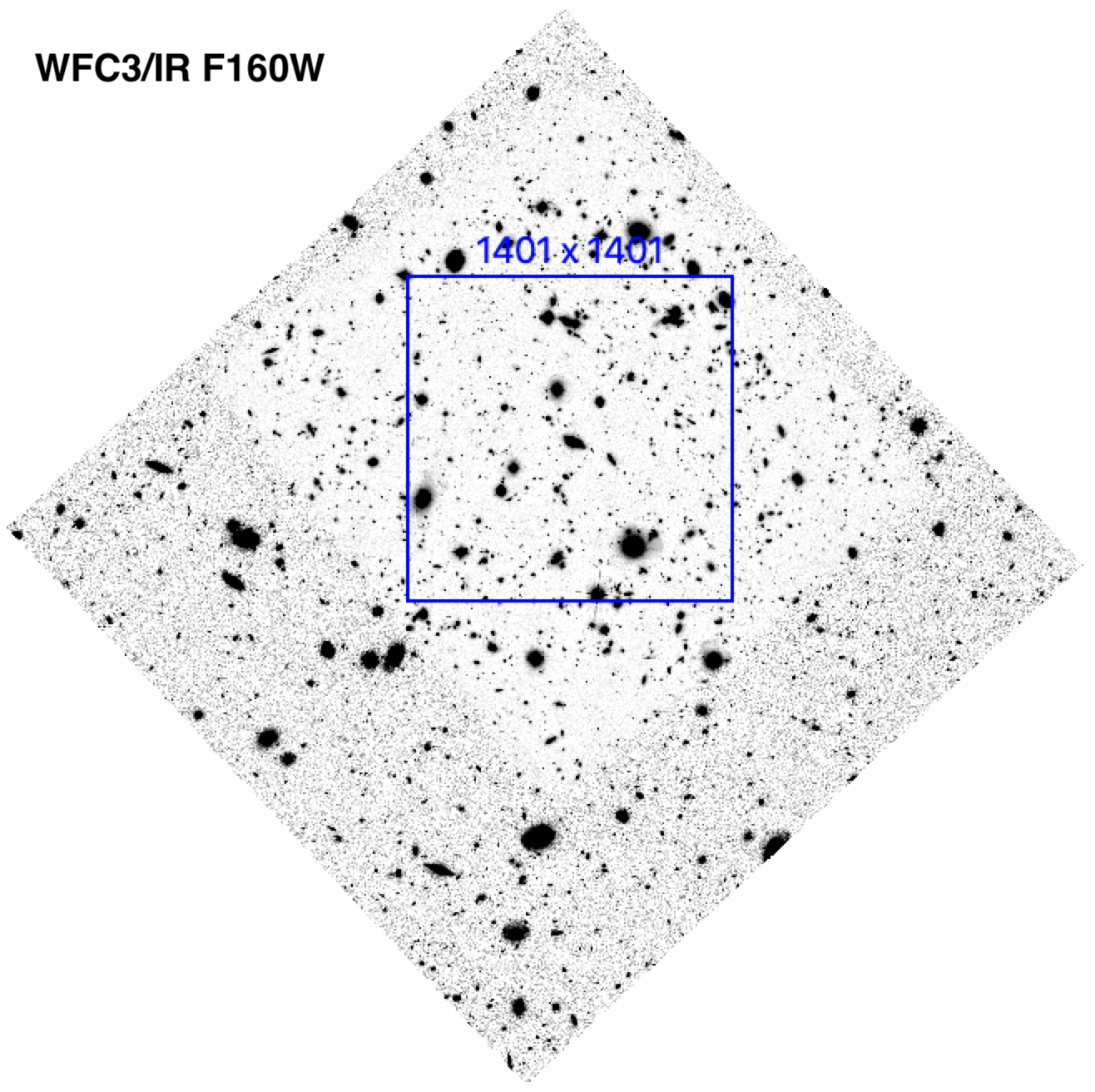}
    \includegraphics[width=0.58\textwidth]{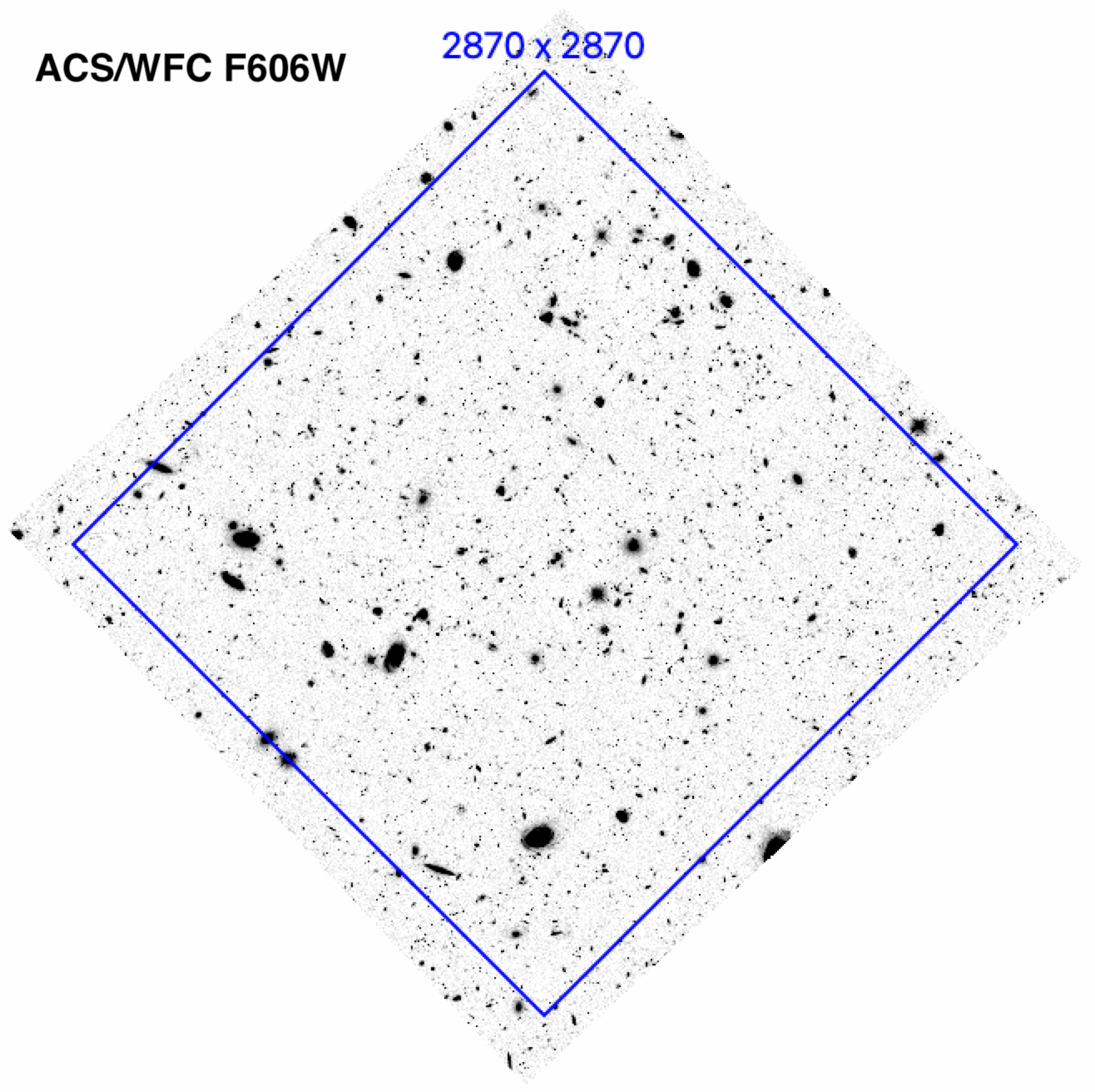}
    \caption{Examples of square image regions excised for analysis from the XDF mosaics before the objects brighter than \mAB\ $\sim$ 24 mag were masked out. (\emph{a}) Exposure times in the WFC3/IR images, such as the F160W image shown here, vary significantly throughout a single mosaic.  We therefore excised a large region with relatively uniform depth in the deepest part of the WFC3/IR mosaics --- a 1401 $\times$ 1401 pixel region centered on pixel (2734,3072) in the XDF mosaics. (\emph{b}) We similarly excised the largest square image regions with relatively uniform depth from the ACS/WFC mosaics, such as the F606W image shown here. Each $\sim$ 2870 $\times$ 2870 pixel region is centered on pixel (2625,2625) in the XDF mosaics.} \label{fig:crop}
\end{figure*}

This is the third paper associated with the SKYSURF project, an Archival \HST\ program (AR~15810; PI: R.~Windhorst) that aims to resolve the direct EBL vs.\ IGL conflict. SKYSURF is addressing the conflict by re-processing \HST\ data in order to measure the sky values, instead of reducing them \citep{rogier2022, tim2022}.

\citet{rogier2022} showed that bright objects with \mAB$<17$ mag contribute $\sim25\%$ of the IGL, while objects 17\cle\mAB\cle22 mag contribute about 50\% and objects fainter than \mAB$=22$ mag contribute 25\% of the IGL. Moreover, objects fainter than \mAB$=24$ contribute less than 10\% of the IGL. Therefore, in order to explain a discrepancy as large as a factor of 3--5 between the direct estimates of the EBL and estimates of the EBL from IGL measurements, one would require as many as 8 times the number of faint galaxies (\mAB$>24$ mag) than are currently detected.

In this work, we study the XDF images that were released by \citet{2013ApJS..209....6I}  to explore how object confusion could affect estimates of the extrapolated galaxy light. We investigate specifically how many times we can replicate the XDF data onto itself, after successive 90\degree\ rotations, before object overlap starts to significantly limit the ability of the deblending algorithm in \texttt{SourceExtractor} to construct complete object catalogs. In \S~\ref{sec:data} we discuss the \HST\ data used in this work and how it was pre-processed. \S~\ref{sec:methods} discusses the use of \texttt{SourceExtractor} on these images. In \S~\ref{sec:results} we show and analyze the \texttt{SourceExtractor} source photometry extracted from the deep field images, and in \S~\ref{sec:conc} the conclusions and implications of these results are laid out. 

\section{Data}\label{sec:data}

It was desired to perform this exercise with data rather than simulations (such as those in \citet{rogier2022} \S\ 4.2) so that galaxy sizes and morphologies could be as realistic as possible. We therefore use the XDF dataset that was released by \citet{2013ApJS..209....6I}. These images contain data obtained with the Advanced Camera for Surveys Wide Field Channel \citep[ACS/WFC,][]{ACS_Handbook} between 2002 and 2013, including new and reprocessed ACS data obtained after the original HUDF to improve its depth, as well as data obtained with the Wide Field Camera 3 InfraRed Channel \citep[WFC3/IR,][]{WFC3_Handbook} between 2009 and 2013. The XDF ACS mosaics are deeper by $\sim 0.12 - 0.2$ magnitudes compared to the original HUDF \citep{2006AJ....132.1729B}, while the XDF WFC3/IR mosaics reach similar depths to the mosaics released from the HUDF12 campaign \citep{Ellis2013,Koekemoer2013} but were reduced completely independently, with both of these releases containing the full set of WFC3/IR data obtained on this field at that time. For consistency with the XDF ACS mosaics, we also make use of the XDF WFC3/IR mosaics in this work. We start with the fully-reduced public XDF mosaics in the F435W, F606W, and F775W broad-band filters, but more filters were used in the analysis (F105W, F125W, F160W). The F140W, F814W, and F850LP filters were omitted from this study because they are shallower than the others used here. The \HST\ data used in this work can be found in MAST: \dataset[10.17909/T97P46]{http://dx.doi.org/10.17909/T97P46}.

To avoid unnecessary complications from the severely declining stacked exposure depth around the edges of the field, we excised image regions from all filters' mosaics that were exactly square and had near-uniform exposure depth. The WFC3/IR mosaics were masked to only include a region with relatively uniform depth in the deepest part of the F105W, F125W, and F160W images --- a 1401$\times$1401 pixel region centered on pixel (2734,\,3072). Fig.~\ref{fig:crop}\emph{a} shows the section that was used compared to the full XDF mosaic. The same square image regions were excised from the associated weight images. 

The ACS/WFC F435W, F606W, and F775W mosaics were masked to excise squares defined to have diagonals measuring 4059 pixels, making their sides approximately 2870 pixels in length (Fig.~\ref{fig:crop}\emph{b}). Each excised $\sim2870\times2870$ pixel region is centered on pixel (2625,\,2625) in the original XDF mosaics.

\begin{figure*}[h!]
    \centering
    \includegraphics[width=\textwidth]{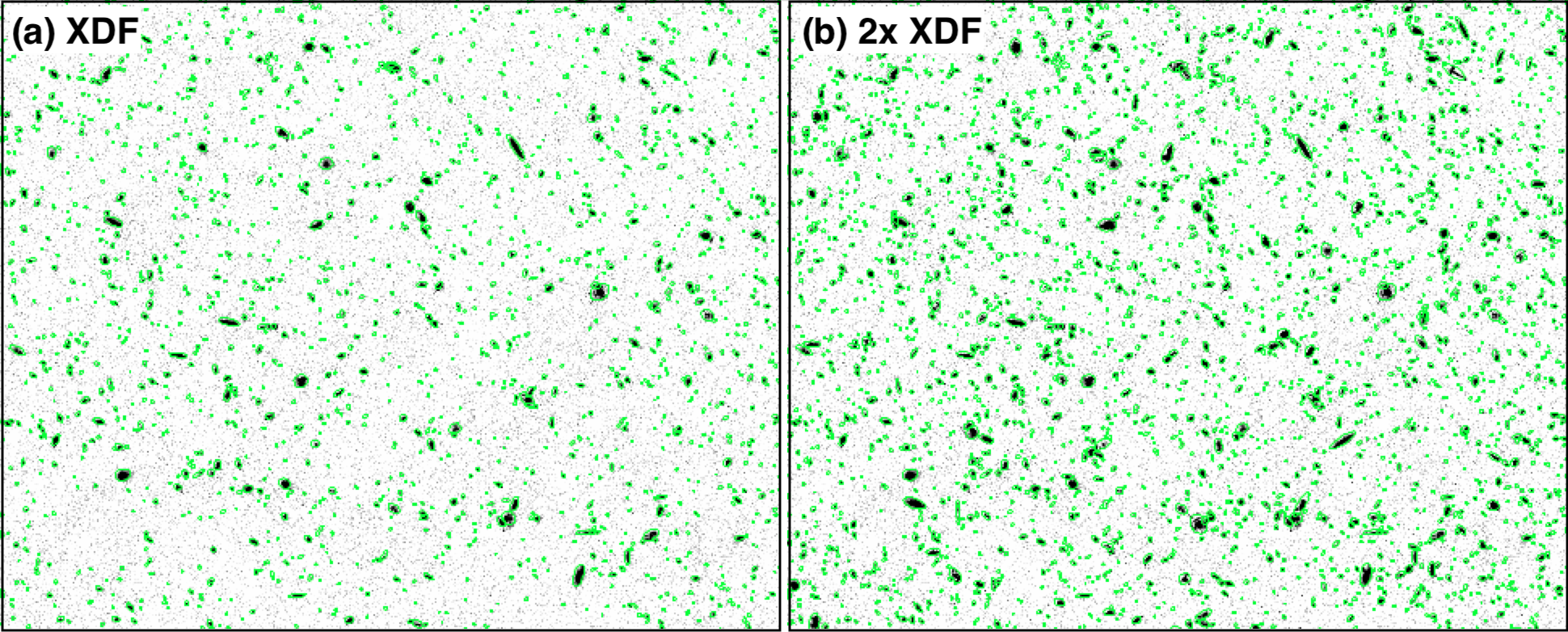} 
    \includegraphics[width=\textwidth]{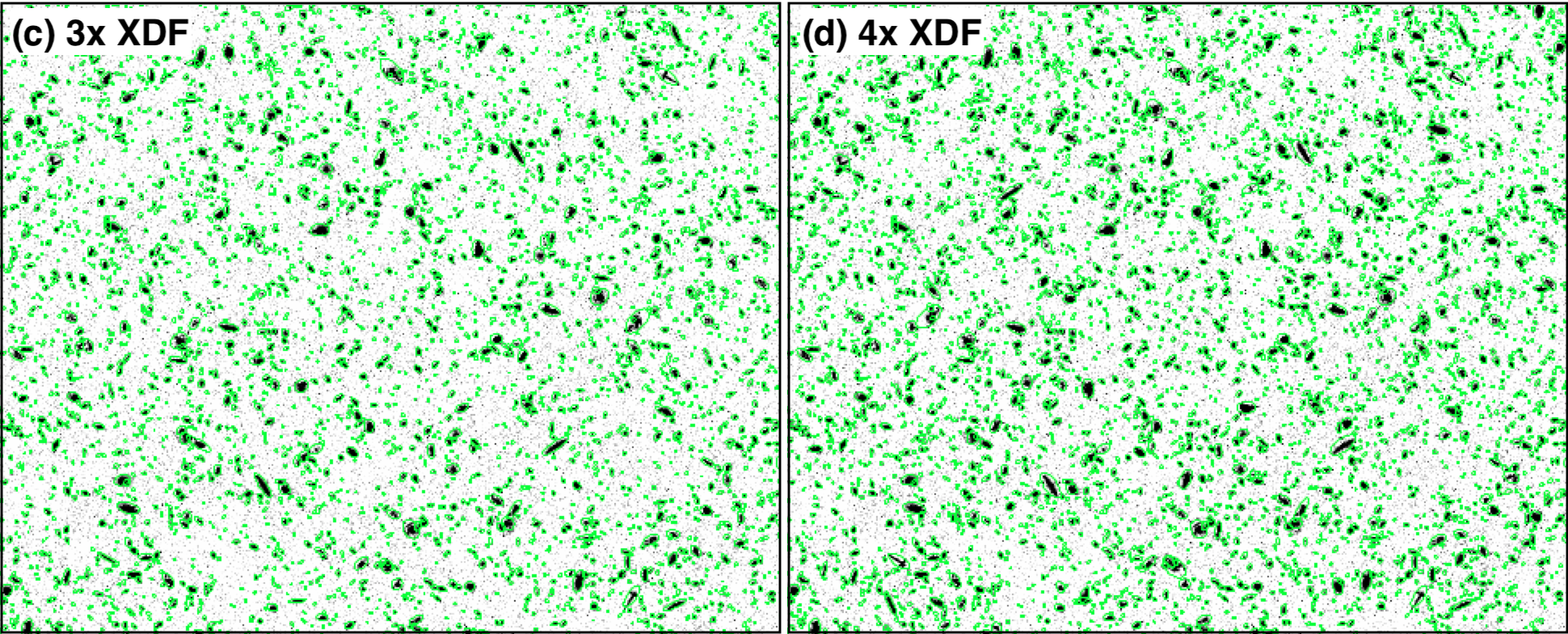}
    \caption{A centered, zoomed comparison of the four images used for the F606W analysis. Each detected object is marked with a green ellipse, the size of which corresponds to $2.5\times$ the profile rms along major and minor axes to account for the Kron factor \citep{Kron1980}.
    (\emph{a}) Original XDF image with objects brighter than \mAB\ $\sim$ 24 mag masked out.
    (\emph{b}) The masked original image + a masked and \emph{once} rotated original image.
    (\emph{c}) The image of panel (b) + a masked and \emph{twice} rotated original image.
    (\emph{d}) The image of panel (c) + a masked and \emph{three times} rotated original image.
    This process increases the surface density of objects brighter than the detection limit by integer factors in each iteration, while leaving the sky unchanged.}
    \label{fig:regcomp}
\end{figure*}

\section{Methods}\label{sec:methods}

\texttt{SourceExtractor} \citep{1996A&AS..117..393B} has been used by astronomers for decades to identify objects in images and extract relevant object properties, including object photometry. We used \texttt{SourceExtractor} along with \texttt{Python} to process and analyze each image following established methods for taking associated weight images into account, and we will use \textsf{MAG\_AUTO} object magnitudes throughout. In this work, we define ``sky" or ``sky values" to mean pixels that were not assigned to part of an object by \texttt{SourceExtractor}.

We wish to investigate only galaxies fainter than \mAB\ = 24 mag because of the possibility that significant numbers of objects in this range are missed due to confusion. For this reason, the objects brighter than \mAB\ = 24 mag were masked from the original XDF data before rotation and addition. To do this, Gaussian distributions were constructed with the sky and root-mean-square (rms) values returned by \texttt{SourceExtractor} for each filter. Because, for example, the 174.4 ks F606W image consists of 286 separate exposures, its noise and weight map are very uniform. Therefore, we drew random sky + noise values from our constructed distribution and placed these in the pixels where the bright objects had been identified by a segmentation map and removed. The segmentation map was created with detection and analysis thresholds of $0.8\times$ the sky rms value (which is $\sim$53\% lower than the $1.5\sigma$ detection values over 4 connected pixels used in the analysis of the final images here) in order to detect and also mask the extended light surrounding bright galaxies. This method eliminated the strong object overlap from large, bright galaxies, and the result is shown in Figure \ref{fig:regcomp} for the masked F606W image and replications thereof. In all, between 3 and 10\% of the pixels in these mosaics were masked.

To prepare for rotation and addition, the original images were run through \texttt{SourceExtractor} with the option to return FITS files with only the objects in them. These object-only FITS images were rotated and co-added to the original mosaics. This minimizes the negative effects of co-adding sky noise.

When the ACS/WFC and WFC3/IR images were drizzled from their native pixels to the 60 milliarcsecond square pixel images (used in this work), the resultant pixels had necessarily correlated noise \citep{Casertano_2001}. Moreover, image rotation by any angle other than $n$ $\times$ {90\degree} and stacking would introduce additional noise due to interpolation between unaligned pixels, whether square or not. We therefore only used image transpositions and inversions to implement the rotations over $n$ $\times$ {90\degree}. Each ``rotated'' image was co-added to the original. This multiplied the number of potentially detectable objects by 2$\times$, 3$\times$, or 4$\times$, respectively. Because the weight maps are inverse variance weights, the sum of the inverses of the original weight maps was inverted to get the replication weight maps. That is, for the 3$\times$ replication, the weight image was constructed from the original XDF weight map and the rotated weight maps:\begin{equation}
    \text{W3x} = \frac{1}{1/\text{W}_0+1/\text{W}_{90}+1/\text{W}_{180}}\ ,
\end{equation}where $\text{W}_{90}$ is the weight image corresponding to a 90\degree\ rotation of the original XDF weight image. \texttt{SourceExtractor} then processed all the images with their appropriate zero points and weight images found on the \href{https://archive.stsci.edu/prepds/xdf/}{XDF Data Release page}; detection and analysis thresholds set to $1.5\sigma$; the minimum number of connected pixels above these thresholds to be a detection set to 4; and minimum contrast for deblending set to 0.02.

\texttt{SourceExtractor} counted the objects in the masked original XDF data, as well as in the replicated data, and returned object catalogs for each. The magnitude and half-light radius columns of the catalog files were used to construct the following figures. Elliptical apertures were generated using the position, size, and orientation columns for each detected object, and imported into \texttt{DS9} \citep{2003ASPC..295..489J} for visual inspection of the objects detected in the respective images. Figure \ref{fig:regcomp} shows examples of these for the F606W filter for all four realizations.

Error calculations in this analysis were performed as follows. For the histograms (left panels in Fig.~\ref{fig:435606775}), the error bars were calculated simply by $\sigma_{\rm bin} = \sqrt{n}$, where $n$ is the number of objects in a particular bin of the original image. For the ratios of the histograms (the right panels in Fig.~\ref{fig:435606775}), the uncertainties are based on their corresponding histograms and derived for a Gaussian error distribution:
\begin{equation}
    \sigma_{\rm bin} =
    \frac{m}{pn}*\sqrt{\left(\frac{\sqrt{m}}{m}\right)^{2} + \left(\frac{\sqrt{pn}}{pn}\right)^{2}}
\end{equation}
where $p$ is the integer by which the original image bins are multiplied (the effective multiple in the transposed and added images), and $m$ is the number of objects in a particular bin of the rotated and co-added images. We note that the rms noise increased by 4--7\% from the initial XDF mosaics to their $4\times$ replications. This can be explained by the wings of some objects going undetected by \texttt{SourceExtractor}, and therefore being included in sky statistics.

We also tested whether using translation rather than rotation affects the results of an analysis like the present one. It turns out that, although the rotation causes slight statistical distortion at the image center, the difference in the results between the two methods is within the counting errors of $\sqrt n$ for each bin, and generally well-within 10\%. As we will see below, this difference between the two replication approaches is much less than the incompleteness levels caused by object overlap that we seek to quantify below. Therefore, we chose to maintain the rotation and addition method, because it dealt well with edge effects (i.e., without cutting additional galaxies in half).

\section{Results}\label{sec:results}

We now wish to examine if the transposed and co-added images indeed contain 2$\times$, 3$\times$, or 4$\times$ as many objects as the original, respectively. \begin{figure*}
    \centering
    \includegraphics[width=\textwidth]{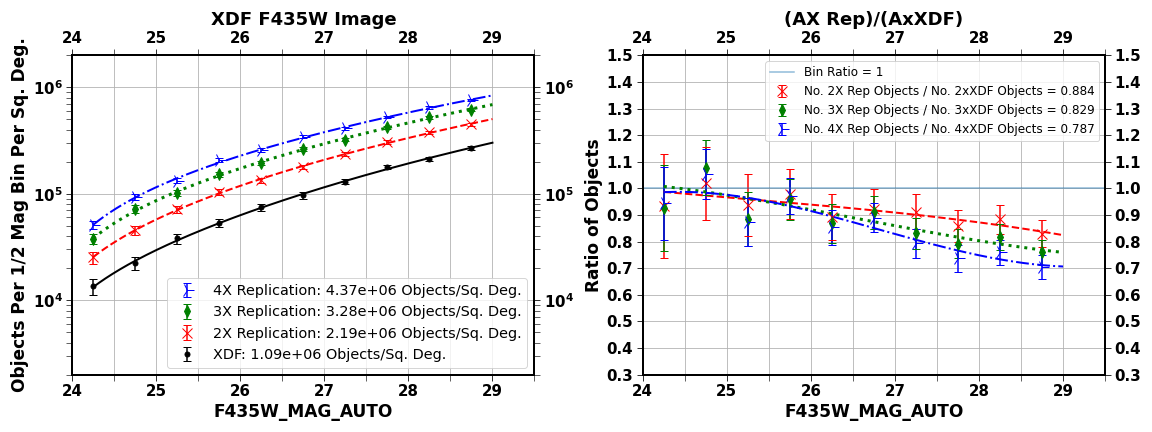}
    \includegraphics[width=\textwidth]{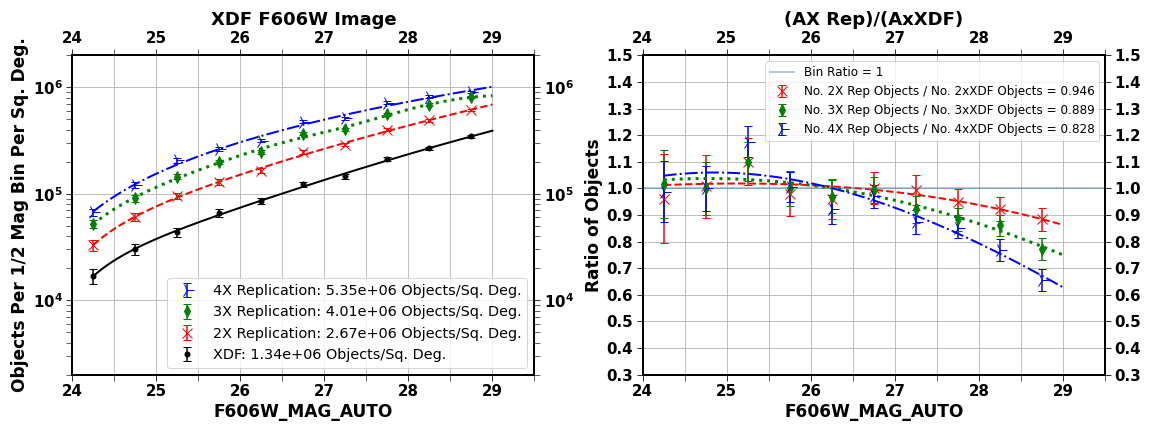}
    \includegraphics[width=\textwidth]{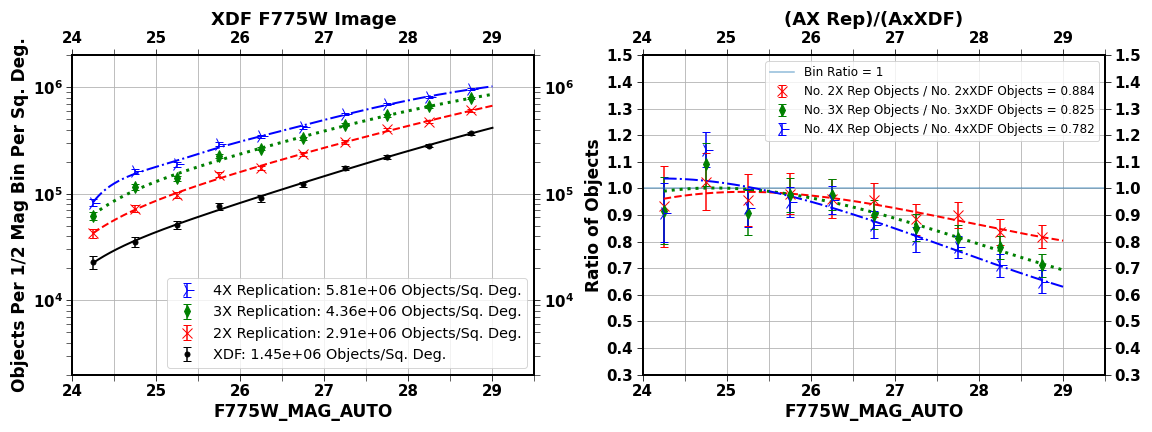}
    \caption{\emph{Left Column:} Histograms of object counts per $\frac{1}{2}$ mag bin per deg$^2$ for the F435W, F606W, and F775W original and replicated images up to their completeness limits. \textbf{AX Rep} refers to the integer A replications of the XDF data, and \textbf{AxXDF} refers to that integer times the original XDF bin counts. Points are located at the centers of their bins. Spline fits show the general trends, and counts are plotted and listed up to their faintest complete magnitudes for the individual filters (e.g., up to \mAB\ = 29 mag for F606W). \emph{Right Column:} The ratios per $\frac{1}{2}$ mag bin of the replicated image bin counts over two, three, and four times the original image bin counts. These curves demonstrate the completeness of each replication. The faint horizontal blue line marks the expected ratio of 1, as a comparison to the actual values.}
    \label{fig:435606775}
\end{figure*} Figure \ref{fig:regcomp} allows a visual comparison of these four images for the F606W filter. The images include (as green ellipses) the apertures associated with each of the objects detected and measured by \texttt{SourceExtractor} in each of the four images. Figure \ref{fig:435606775} shows the results from the above methods as applied to XDF F435W, F606W, and F775W images in the HUDF. The left column of Figure \ref{fig:435606775} shows the differential object counts as a function of \mAB\ for each of the four realizations, as indicated by the four different colors. These are the raw data before correcting for the fact that each of the realizations has 2$\times$, 3$\times$, or 4$\times$ as many objects as the original image (shown in black). Some objects sufficiently overlap neighbors such that \texttt{SourceExtractor} deemed them to be a single object, causing the count slopes to slightly change between each iteration. While the (black) input catalog had no objects brighter than \mAB\ = 24 mag, object overlap even resulted in some objects being slightly brighter than this limit (these points are present but excluded in the figures here). The panels in the right column of Figure \ref{fig:435606775} show the ratios of the counts in the left panels and the original counts (in black), but multiplied by the proper integer. Therefore, these ``completeness functions'' in the right panels show the decreasing trend in completeness as each replication becomes more and more dense.

\begin{figure*}
    \centering
    \includegraphics[width=\textwidth]{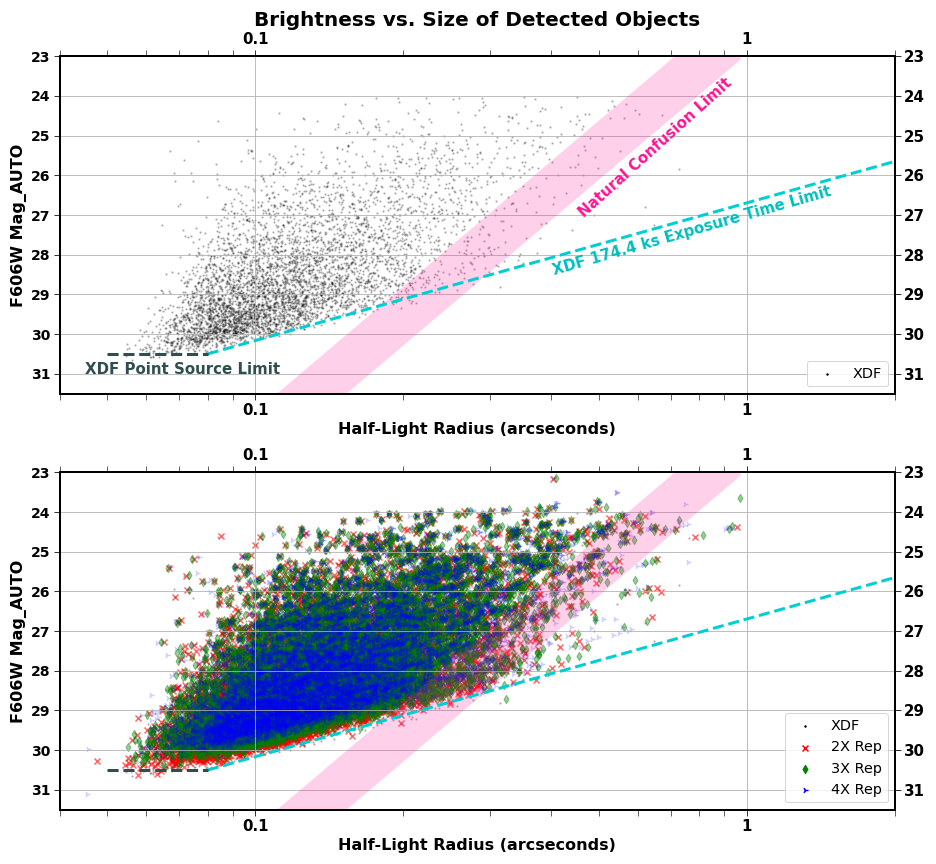}
    \caption{Magnitude vs. Half-Light Radius plot for the F606W original and replicated images. The black points are the original objects, and the other colors are the 2$\times$, 3$\times$, and 4$\times$ XDF replications' objects. Each successive color has more transparent markers, and there are more blue marks than green, more green than red, and more red than black. This figure was modeled after Figure 3 in \citet{Windhorst2008} so features could be marked and compared. The pink band represents the natural confusion limit with a range between 1/25 and 1/50 objects per beam \citep[see e.g.][]{Serjeant_1997, Silva_2004}. The cyan dashed curve represents the 592-hour surface brightness limit of XDF data. Its shape has been changed from \citet{Windhorst2008} as the curve is only intended to draw the eye to the trend and the cutoff of the objects, demonstrating the exposure time and resolution limits to observing these objects. The dark grey dashed line represents the limit below which point sources with fainter magnitudes cannot be resolved, as they blend in with the sky distribution. All curves have been shifted in magnitude to account for the $m_{Vega}$ to \mAB\ conversion. This figure demonstrates that natural confusion (pink band) is an important completeness constraint for bright, large objects that make up most of the IGL (\mAB\cle28.5 mag). It also demonstrates that the three main limits on the original HUDF sample remain the the same for the replicated samples, validating our methods of replication, namely the point source detection limit (horizontal grey dashed line), the SB-limit (slanted cyan dashed line) and the natural confusion limit due to statistical object overlap (pink band).}
    \label{fig:elephant}
\end{figure*}

Figure \ref{fig:elephant} shows a magnitude-radius plot for all four F606W samples, which allows us to compare these samples according to their angular size and magnitude. The nearly horizontal limit that ends at \mAB\ $\simeq$ 30.5 mag is the point-source sensitivity limit. The slanted cyan line that extends into the upper right of Figure \ref{fig:elephant} is the surface brightness limit that indicates how well larger but dim objects can be resolved in the F606W image with its 174.4\,ks total exposure time. The slanted pink band indicates the approximate natural confusion limit of 1 object / 25 beams to 1 object / 50 beams, to the right of which objects cannot be easily detected because they start to significantly overlap in the \texttt{SourceExtractor} segmentation maps \citep[e.g.,][]{Windhorst2008}. Note that in the concept of natural confusion of resolved objects, ``beam'' refers to the average object area at that AB-magnitude level, or approximately $\pi r_e^2$, where $r_e$ is the median object half-light radius at that flux level in Figure \ref{fig:elephant}.
\begin{table}[tb]
\movetableright=-0.55in
\setlength{\tabcolsep}{1.5pt}
\begin{tabular}{lcccccc}
\hline\hline\\[-10pt]
Sample      & \multicolumn{6}{c}{\hrulefill\ Filter \hrulefill}\\ 
            & F435W$^\star$ & F606W$^\star$ & F775W$^\star$ & F105W$^{\dag}$ & F125W$^{\dag}$ & F160W$^{\dag}$ \\[3pt]
\hline\\[-10pt]
HUDF09 & 0.237 & 0.228 & 0.222 & 0.209 & 0.226 & 0.209 \\
XDF & 0.241 & 0.248 & 0.238 & 0.249 & 0.249 & 0.232 \\
2$\times$XDF & 0.223 & 0.230 & 0.218 & 0.211 & 0.219 & 0.206 \\
3$\times$XDF & 0.213 & 0.213 & 0.203 & 0.178 & 0.191 & 0.163 \\
4$\times$XDF & 0.206 & 0.190 & 0.191 & 0.149 & 0.173 & 0.129 \\
\hline\\[-4pt]
\end{tabular}
\begin{minipage}{0.475\textwidth}{\small
$^\star$ indicates \mAB\ 25-29 slope

$^\dag$ indicates \mAB\ 25-28 slope}
\end{minipage}
\caption{Slopes (in dex per 0.5 magnitude) of the samples in this work as compared to those measured by \citet{2011ApJS..193...27W} in the HUDF09 field \citep{hudf09}. XDF count slopes in this work are generally larger than those in \citet{2011ApJS..193...27W} Figure 12, which is likely due to a higher percentage of completeness in the XDF data. The replicated slopes in this work are also expected to be flatter with each successive replication of the XDF data, as is indeed observed.\label{tab:slopes}}
\end{table}

Comparing slopes in the left column of Figure \ref{fig:435606775} to Figure 11 of \citet{2011ApJS..193...27W}, we find that those in the present work are generally flatter (see Table~\ref{tab:slopes}). Because only fainter magnitudes were used here, this is expected. First, the slopes are expected to flatten due to incompleteness at the fainter magnitudes. Second, the same flattening trend is also seen between iterations of replicating the images due to the increasing severity of the simulated natural confusion resulting in increasing incompleteness. For the corresponding results for the WFC3/IR F105W, F125W, and F160W filters, we refer the reader to the Appendix and Table \ref{tab:slopes}. These results are consistent with the findings already presented, and all six filters will be discussed in \S\ \ref{sec:conc}.

It is worth noting that the exercises in this paper were initially performed on un-masked XDF data. Masking brighter objects was later desired to narrow the focus of this work toward the faint objects that are of importance in this discussion. Masking the brighter objects before rotation and addition and comparing to un-masked replications showed that the un-masked analysis had $\sim$5-30\% lower total relative number counts (right column of Figures \ref{fig:435606775} and \ref{fig:105125160}), meaning fewer objects were recovered in the un-masked images than in the masked ones. When these object counts were weighted by their luminosities and summed to get faint-end IGL calculations, the masked IGL values for each filter for objects with 24\cle\mAB\cle29 mag were on average 10\% lower than those from the un-masked images. This is due to the generally-larger number of objects in the un-masked image. Therefore, our completeness curves and summed values in each \HST\ filter, in the right panels of Figures \ref{fig:435606775} and \ref{fig:105125160}, are upper limits.

\section{Discussion \& Conclusion}\label{sec:conc}

To quantify the effect of possible missing faint galaxies on the total IGL, we calculated and summed the IGL fluxes in each filter and for each replication. Then, using the fact that galaxies with 24\cle\mAB\cle29 mag make up at most 10\% of the IGL \citep{rogier2022}, we used the equation below to calculate the percent change in the IGL resulting from the $2\times,$ $3\times$, and $4\times$ replications:
\begin{equation}
\Delta\text{IGL}_{TOT}(n) = \frac{F_n - F_1}{10F_1}
\end{equation}
where $F_n$ is the integral of the flux for the $n$th replication, and $F_1$ is the integral of the flux from the original XDF data in that given filter. $F_1$ is multiplied by 10 simply because it represents 10\% of the IGL on its own. The denominator therefore represents 100\% of the IGL. The results of this analysis show that, in all filters, doubling the number of detectable objects with 24\cle\mAB\cle29\ mag increases the total IGL by 8-10\%, tripling the number increases the IGL by 16-20\%, and quadrupling the number increases it by 24-30\%. For the F606W filter, this means that the missing IGL due to natural confusion increases to $30\%$ as the galaxy number counts increase from $\sim1.3\times 10^6$ to $\sim5.3\times 10^6$ objects per square degree. 

To address how many galaxies could be missing from the current XDF data, we extrapolated the linear trend in relative source counts to a $1\times$ value for the three ACS/WFC filters. This calculation yields estimated total completeness percentages at faint magnitudes of 93\% for F435W filter, 99\% for the F606W filter, and 94\% for the F775W filter.

\begin{figure}[htp!]
    \centering
    \includegraphics[width=0.88\textwidth]{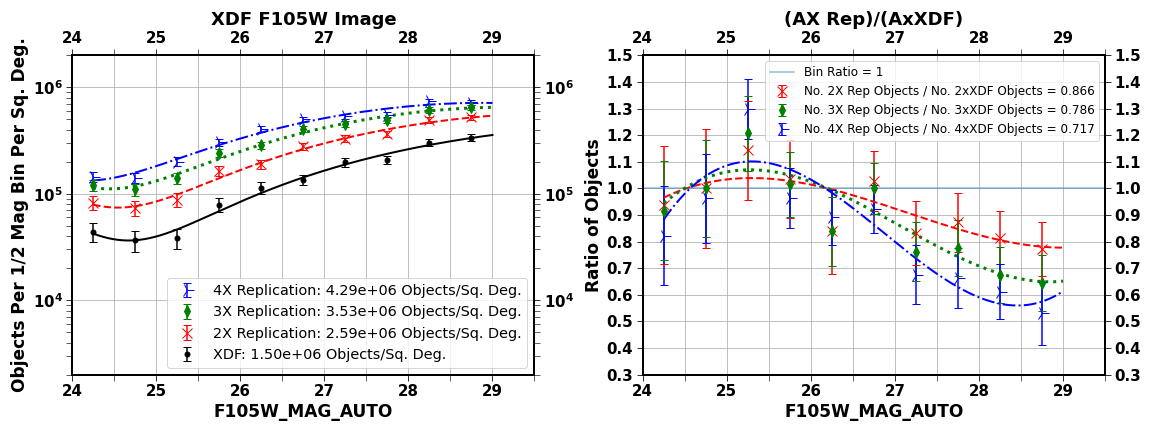}
    \includegraphics[width=0.88\textwidth]{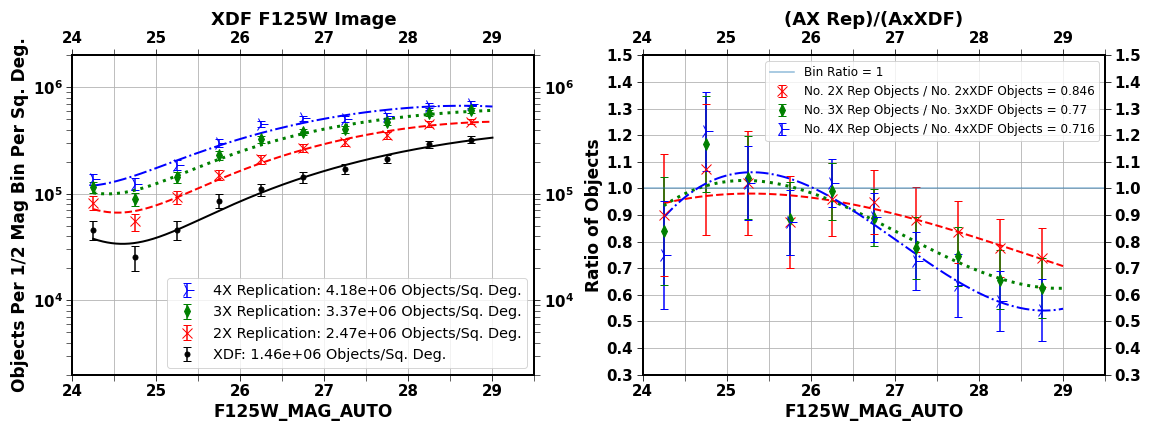}
    \includegraphics[width=0.88\textwidth]{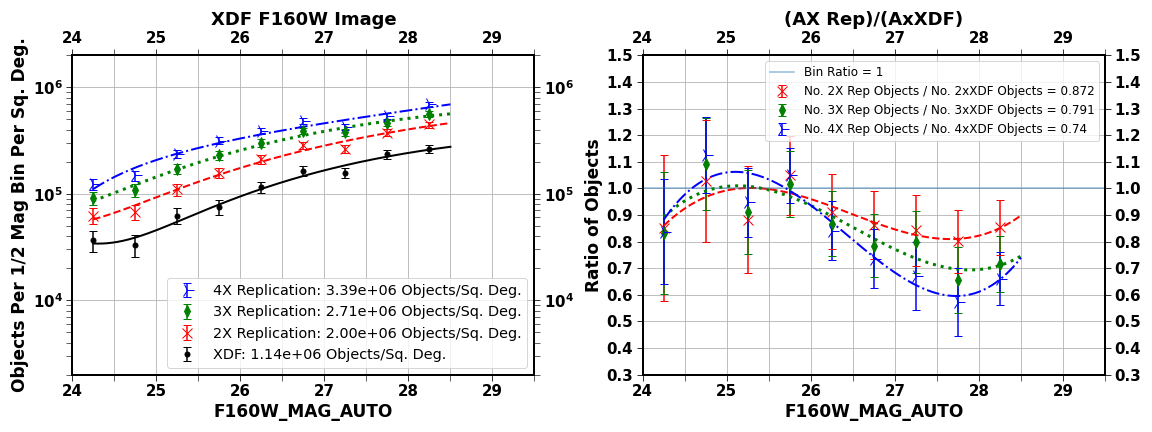}
    \caption{\emph{Left Column:} Histograms of object counts per $\frac{1}{2}$ mag bin per deg$^2$ for the F105W, F125W, and F160W original and replicated images up to their completeness limits. \textbf{AX Rep} refers to the integer A replications of the XDF data, and \textbf{AxXDF} refers to that integer times the original XDF bin counts. See Figure \ref{fig:435606775} for the ACS/WFC filters. The values in the legend are up to the completeness limit in each filter and replication. \emph{Right Column:} The ratios per $\frac{1}{2}$ mag bin of the replicated image bin counts over two, three, and four times the original image bin counts. These curves demonstrate the completeness of each replication and how it varies with magnitude.}\label{fig:105125160}
\end{figure}

Based on this analysis, it does not appear likely that missing faint galaxies explain the optical EBL discrepancy. The completeness limits of the WFC3/IR images analyzed here (see the Appendix) were slightly lower than that of the ACS/WFC images. This is likely due to the larger FWHM at the wavelengths of the WFC3/IR filters. See \href{http://svo2.cab.inta-csic.es/theory/fps/index.php?mode=browse&gname=HST&gname2=WFC3_IR&asttype=}{SVO Filter Profile Service} and \citet[their Table~2]{2011ApJS..193...27W} for the FWHMs and more info on these \HST\ filters. The WFC3/IR filters were analyzed in this work mostly to ensure that the analysis was consistent, and to determine there was no major wavelength or FWHM dependence in our results. The quoted fractions in the legends of Figures \ref{fig:435606775} and \ref{fig:105125160} are the result of the total number of objects with 24\cle\mAB\cle29 recovered in the denser realizations compared to the original after appropriate normalization. One can still analyze completeness at specific magnitudes based on these results. For example, in most filters the downward trend in the ratio points does not begin until \mAB\ $\sim$ 27 mag, and most filters were at least 50\% complete up to \mAB$=29$ mag, even in the $4\times$ realizations. In other HUDF works, object counts are also done in other filters, but we confined ourselves to the six deepest HUDF filters, which exclude F814W, F850LP, F140W, and the 3 UV filters of \citet{teplitz2013}.

It is clear that significant numbers of faint galaxies are still missed or blended with other objects in this experiment. This is expected because of the large number of galaxies that are being added to the original, especially in the 3$\times$ and 4$\times$ replications. Natural confusion and surface brightness limitations cause the incompleteness of the sample to be noticeable for sizes bigger than 0\farcs15--0\farcs2, and magnitudes \mAB\cge 29 mag. Even so, our conclusions are two-fold. If there existed 4 times more faint galaxies in our Universe than what we observe today, then: \textbf{(1)} we would only miss at most 30\% of them (i.e., the highest, $4\times$-replicated relative number count deficit in Figure \ref{fig:105125160}), and so galaxy catalogs would contain many more galaxies than are present in the actual XDF data; and \textbf{(2)} not nearly enough IGL would be missed due to object overlap in this sample (at most 30\% of the total) to account for the 100\% IGL-EBL discrepancy.

Further studies of truly hierarchical simulations could be done to find out how many, and what kind, of galaxies are missed on a portion of the sky when their total surface density changes. The SKYSURF project \citep{rogier2022,tim2022} is working to resolve the EBL discrepancy by reprocessing \HST\ data to include and measure the sky surface brightness for both EBL and diffuse Zodiacal light calculations.

\section*{Acknowledgments}
We thank the anonymous referee for insightful comments that helped improve this paper. We also thank the entire SKYSURF team for their interest, comments, and discussion about this work; especially, we thank Rosalia O'Brien for her support and helpful feedback. Finally, we thank Alex Van Engelen and his research group at ASU for their insight and perspective. This project is based on observations made with the NASA/ESA \emph{Hubble Space Telescope} and obtained from the Hubble Legacy Archive, which is a collaboration between the Space Telescope Science Institute (STScI/NASA), the Space Telescope European Coordinating Facility (ST-ECF/ESA), and the Canadian
Astronomy Data Centre (CADC/NRC/CSA). We acknowledge support for \HST\ program AR-15810 provided by NASA through a grant from STScI, which is operated by the Association of Universities for Research in Astronomy, Incorporated, under NASA contract NAS5-26555. 

\software{\texttt{SourceExtractor} \citep{1996A&AS..117..393B}}

\bibliography{BIB} \section*{Appendix}\label{sec:Appendix}
Below are the equivalent plots to those in Figure \ref{fig:435606775} for the remaining XDF filters mentioned above. Trends are similar, though for the WFC3/IR filters the total percent recovery of object counts are $\sim$10-20\% lower than those in the ACS/WFC filters due to the filters' higher FWHMs, as discussed in \S\ \ref{sec:conc}.
\end{document}